\documentclass[lettersize,journal]{IEEEtran}
\usepackage{amsmath,amsfonts}
\usepackage{algpseudocode}
\usepackage{algorithm}
\usepackage{array}
\usepackage[caption=false,font=normalsize,labelfont=sf,textfont=sf]{subfig}
\usepackage{textcomp}
\usepackage{stfloats}
\usepackage{url}
\usepackage{verbatim}
\usepackage{graphicx}
\usepackage{cite}
\usepackage{xcolor}
\begin{document}

\title{Structure Aware Image Downscaling}


\author{G B Kevin Arjun,  Suvrojit Mitra, and Sanjay Ghosh, \IEEEmembership{Senior Member, IEEE}
\thanks{This research is supported by the Faculty Start-up Research Grant (FSRG), IIT Kharagpur, awarded to Dr. Sanjay Ghosh.}
\thanks{G B Kevin Arjun,  Suvrojit Mitra, and Sanjay Ghosh are with Department of Electrical Engineering, Indian Institute of Technology Kharagpur, WB 721302, India (e-mail: \texttt{sanjay.ghosh@ee.iitkgp.ac.in}).}
}

\markboth{Under-Review}%
{Shell \MakeLowercase{\textit{Arjun et al.}}: A Sample Article Using IEEEtran.cls for IEEE Journals}



\maketitle

\begin{abstract}
Image downscaling is one of the key operations in recent display technology and visualization tools. By this process, the dimension of an image is reduced, aiming to preserve structural integrity and visual fidelity. In this paper, we propose a new image downscaling method which is built on the core ideas of image filtering and edge detection. In particular,  we present a structure-informed downscaling algorithm that maintains fine details through edge-aware processing. The proposed method comprises three steps: (i) edge map computation, (ii) edge-guided interpolation, and (iii) texture enhancement. To faithfully retain the strong structures in an image, we first compute the edge maps by applying an efficient edge detection operator. This is followed by an edge-guided interpolation to preserve fine details after resizing. Finally, we fuse local texture enriched component of the original image to the interpolated one to restore high-frequency information. By integrating edge information with adaptive filtering, our approach effectively minimizes artifacts while retaining crucial image features. To demonstrate the effective downscaling capability of our proposed method, we validate on four datasets: DIV2K, BSD100, Urban100, and RealSR. For downscaling by $4\times$, our method could achieve as high as $39.07$ dB PSNR on the DIV2K dataset and $38.71$ dB on the RealSR dataset. Extensive experimental results confirm that the proposed image downscaling method is capable of achieving superior performance in terms of both visual quality and performance metrics with reference to recent methods. Most importantly, the downscaled images by our method do not suffer from edge blurring and texture loss, unlike many existing ones. 
\\ \textcolor{blue}{Our codes will be publicly available upon publication.}
\end{abstract}


\section{Introduction}
\IEEEPARstart{T}{he} advancements in display technology have led to increased use of high-resolution images, which are often downscaled for display, storage, and transmission optimization. Image downscaling is crucial for applications such as compression \cite{liu2019downscaling}, remote-sensing \cite{sdraka2022deep}, video processing \cite{hwang1997high,tan2004fast,kim2018exploiting}, rendering \cite{low2017multi}, and streaming \cite{liu2014content}
with the challenge of preserving structural and textural details.
A study on the effect of canny edge detection on downscaled images suggests that an efficient images downscaling method helps to retain the crucial edges which were present in the original image (before performing the resizing operation) \cite{kim2020study, zha2023conditional}. 
Traditional methods like bilinear and bicubic interpolation \cite{keys1981cubic} can introduce blurring and aliasing. Lanczos filters \cite{lanczos1964interpolation} sharpen images but can create ringing artifacts, especially at lower resolutions. L0-regularization \cite{liu2018l0} has demonstrated improvements in preserving structural details during downscaling, while co-occurrence learning \cite{ghosh2023image} has captured better texture information via co-occurrence frequency.
Edge-aware methods \cite{canny1986edge, marr1980theory} use edge information for better interpolation but still face difficulties with complex textures. 
Edge-guided interpolation \cite{xu2009edge} enhances structure preservation but struggles with texture consistency. Adaptive downscaling methods \cite{kopf2013content, weber2016rapid} attempt to balance detail preservation and efficiency but may not generalize well across diverse scaling factors.

Deep learning methods, including CNNs and autoencoders \cite{dong2016image, vincent2008denoising}, learn mappings from high- to low-resolution images but require large datasets and significant resources. Techniques like residual learning-based denoising CNNs \cite{zhang2018beyond} improve performance but still struggle with texture preservation and non-integer scaling. GAN-based approaches such as SRGAN \cite{ledig2017photo} generate visually appealing results but may introduce artifacts, especially at lower resolutions. Multi-scale networks, including pixel-level interpolation \cite{yang2007downscaling}, enhance quality across different scales, though texture and edge preservation remain challenging.
Among recent advancements,  Kim et al. introduced an autoencoder framework that jointly learns downscaling and upscaling networks, optimizing performance for restoration tasks \cite{kim2018task}. Xing et al. developed a versatile network capable of handling arbitrary scale factors in a reversible manner, enhancing real-world applicability \cite{xing2023scale}. Authors in \cite{zhang2023adaptive} proposed a method for simultaneously optimizing rate, accuracy, and latency, offering a practical solution for real-time image processing. Most recently, Liang et al. introduced IDA-RD, a novel metric for quantitatively evaluating downscaling algorithms \cite{liang2024deep}.
While highlighting the importance of an efficient image downscaling technique, authors in \cite{liang2024deep} proposed a new measure to quantitatively evaluate image downscaling algorithms based on the idea that downscaling and super-resolution (SR) can be viewed as the encoding and decoding processes in the rate-distortion model. 

\begin{figure*}[h]
    \centering
    \includegraphics[width=0.8\textwidth]{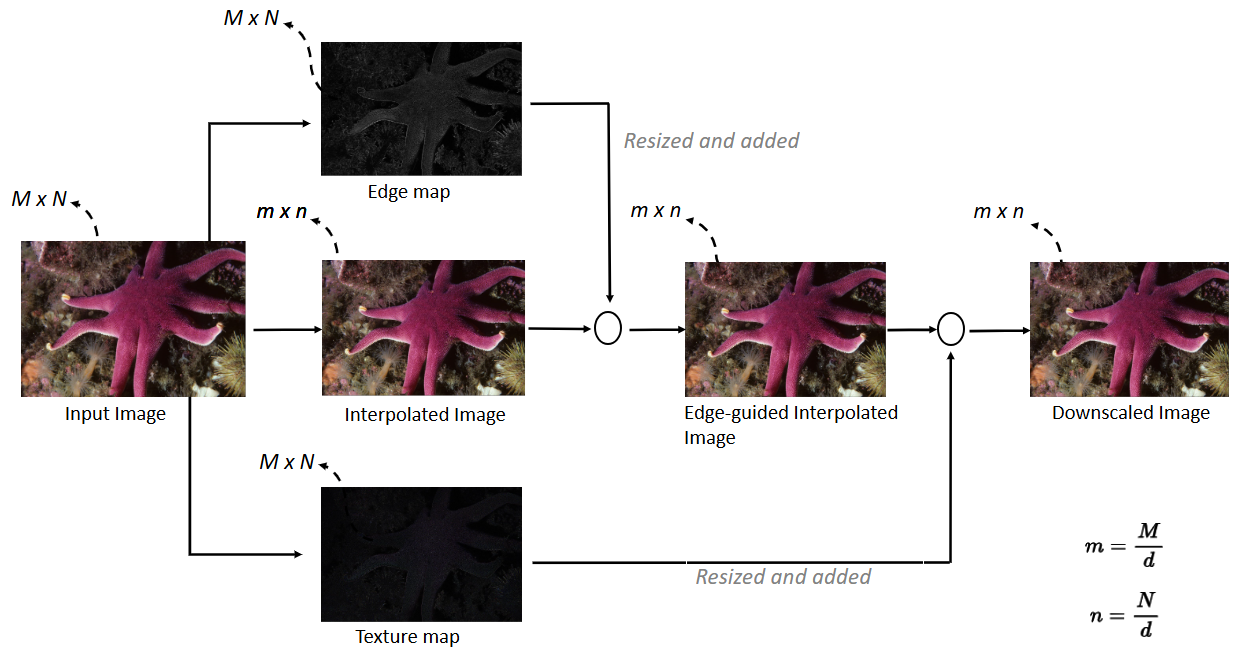}  
    \caption{The input image is first downsampled using interpolation to obtain a coarse low-resolution version. Edge and texture maps are extracted from the original high-resolution image, resized, and sequentially added to enhance structural and textural details. This results in an edge-guided interpolated image, which is further refined with texture enhancement to produce the final downscaled image. The process preserves important visual features during resolution reduction. }
    \label{fig:schematic}
\end{figure*}

\textbf{Contributions:}
In this paper, we propose a structure-informed downscaling algorithm that combines edge-guided interpolation and texture expansion to preserve fine details. First, we construct an edge map from the input image (to be downscaled) using the Sobel operator to maintain structural integrity. Next, we perform a modified bicubic interpolation to adjust pixel intensity based on edge strength to reduce blurring. At the third stage, we apply a Laplacian-based filtering process to build a texture map of the input image. A schematic of our proposed method is shown in Figure \ref{fig:schematic}. With elaborate experiments, we show that the proposed downscaling method outperforms existing methods in preserving structure and minimizing artifacts. The proposed method can be extended for downscaling an image by a non-integer scaling factors;
which would be suitable for many real-world applications.
The main contributions of this
work are summarized as follows:

\begin{enumerate}
    \item To the best of our knowledge, our proposed method is the
most efficient kernel-based image downscaler.  The proposed method can also support all downscaling factors with low complexity that can be used in modern display devices. In particular, computational simplicity is the key aspect of the proposed method. The phenomenon makes the best suited for real-time implementation on low-power settings.

\item  By performing fusion of strong structures and high-frequency texture during the downscaling process, we achieve output that does not suffer from blurring and loss of fine details. Experimental results indicate a new state-of-the-art classical downscaling method for a variety of images such as natural, cartoon, texture, text, etc.

\item We validate our method on four standard datasets for different downscaling and achieve state-of-the-art results. Importantly, our proposed classical (non-deep) method outperforms state-of-the-art deep learning methods. The computational simplicity of our method makes it suitable for hardware and real-time applications. 


\end{enumerate}

\textbf{Organization:}
The rest of the paper is organized as follows. We present a
review of the literature on image downscaling in Section \ref{sec:related}. The proposed downscaling algorithm is described in Section \ref{sec:method}. We report several experiments, exhibiting very competitive results in Section \ref{sec:results}. To demonstrate the effectiveness of our proposal, we performed
exhaustive downscaling experiments with a wide range of image classes and
different downscaling factor. Finally, we conclude in Section \ref{sec:conclusion}.
\section{Related Work}
\label{sec:related}
Image downscaling is a fundamental process in image processing that aims to reduce the spatial resolution of images while maintaining perceptual quality. It plays a vital role in applications such as image compression, visualization on low-resolution displays, and transmission over limited-bandwidth networks. The key challenge is to preserve important structural and textural information—such as edges and contours—while eliminating redundant data. A wide range of existing methods are evolving from traditional interpolation-based methods to modern deep learning models.

\subsection{Interpolation-Based Approaches}

Traditional downscaling methods primarily rely on interpolation schemes with fixed kernels. Bicubic interpolation~\cite{keys1981cubic} is one of the most widely used techniques due to its simplicity and relatively good visual performance. It computes the output pixel value based on a weighted average of surrounding pixels, providing smoother results than nearest-neighbor or bilinear interpolation. However, it often introduces noticeable blurring in high-frequency regions, failing to preserve edge sharpness.
Lanczos resampling~\cite{lanczos1964interpolation} improves upon bicubic interpolation by using a windowed sinc function as the kernel. This method offers better preservation of image sharpness and frequency components but comes at the cost of increased computational complexity, making it less suitable for real-time applications.
{Authors in \cite{occorsio2023image} introduced a downscaling based on non-uniform sampling grids and employs the filtered \textit{de la Vall´ee Poussin}
type polynomial interpolation. Another kernel approximation based work was presented in \cite{occorsio2022lagrange}. The idea was to approximate the image at the desired size by means of global interpolation processes based on (tensor product) Chebyshev grids of I kind. A recent method uses edge-guided interpolation followed by adaptive bilateral filtering \cite{park2020edge}.}


To address the limitations of spatial-only interpolation methods, edge-aware filtering techniques have been employed. 
%
Saliency-guided downscaling methods~\cite{kopf2013content} enhance visual fidelity by allocating more sampling resources to visually important regions. These methods typically compute an importance map using visual saliency cues, which then guide the resampling process to focus on content-rich areas while discarding less significant regions.




\subsection{Optimization-Based Techniques}

Optimization-driven approaches formulate downscaling as an inverse problem and seek to minimize a loss function that measures perceptual similarity between the high-resolution input and the downscaled output. A prominent example is the use of Structural Similarity Index (SSIM)~\cite{wang2004image} as an objective function to better model human visual perception.
Liu et al.~\cite{liu2018l0} introduced an L0 gradient minimization method that encourages sparsity in the image gradients. By suppressing minor variations while retaining significant edge structures, this approach effectively preserves important image features during downscaling. A solution to the $L_0$-regularized optimization problem based on deep unsupervised learning was proposed in \cite{yang2024parameterized}. The $L_0$ norm is first approximated using $L_1$ norm of with varying parameters, which are then solved via training a single fully convolutional network. The authors in \cite{pan2022towards} proposed a joint optimization process that is able to learn arbitrary downscaling and arbitrary upscaling simultaneously. 



\subsection{Deep Learning-Based Models}

Recent advances in deep learning have significantly transformed the image downscaling landscape. Convolutional Neural Networks (CNNs) have shown exceptional capability in learning complex mappings between high-resolution and low-resolution images. Vasileiou et al.~\cite{vasileiou2022efficient} demonstrated the effectiveness of CNN-based downscaling for enhancing resolution in millimeter-wave radar imagery.
Dong et al.’s SRCNN model~\cite{dong2016image, dong2016fast}, although originally developed for super-resolution, has been adapted for downscaling by reversing the training objective. The model learns to map high-resolution patches to their low-resolution counterparts, providing an end-to-end solution that outperforms traditional methods in perceptual quality. 
Hybrid techniques such as those proposed by Kim et al.~\cite{kim2018task} combine classical interpolation (e.g., bicubic) with deep learning refinement. These methods strike a balance between interpretability and performance by first downscaling with conventional techniques and then using CNNs to enhance perceptual features.
{A resampler network, introduced in \cite{sun2020learned}, generates content adaptive image resampling kernels that are applied to the original HR input to generate pixels in the downscaled image. 
Chen et al. \cite{chen2022convolutional} proposed a convolutional block design for fractional downsampling factor, which is of great importance in many practical image and video processing applications. Zhang et al. \cite{zhang2022enhancing} proposed a low-complex invertible image downscaling model by using latent variable within their architecture. Another invertible network which supports 360 degree image downscaling was introduced in \cite{guo2023dinn360}. Recently, a  compression-aware image rescaling method was introduced in \cite{li2025lightweight}.}

{Many researcher are actively working on video downscaling methods \cite{chen2024learned, xiang2022learning, cho2021learning}.
Chen et al. \cite{chen2024learned} proposed a progressive residual learning network architecture for video downsampling in streaming. It was found \cite{xiang2022learning} that improper spatio-temporal downsampling applied on videos can cause aliasing issues. To overcome this issue, authors in \cite{xiang2022learning} propose a neural network framework that jointly learns spatio-temporal downsampling and upsampling. Cho et al. \cite{cho2021learning} proposed a kernel-learning-based image downscaler which supports arbitrary downscaling factors using simple linear mapping.}
Encoder-decoder architectures are widely used in this domain, where the encoder extracts a compact representation of the image and the decoder reconstructs a perceptually consistent downscaled version. Generative Adversarial Networks (GANs) have also been utilized to improve realism. For example, Zhang et al.~\cite{zhang2018beyond} introduced a GAN-based framework where the generator learns to produce visually plausible downscaled images. 

\section{Proposed Method}
\label{sec:method}
The proposed algorithm consists of three major steps:  edge map computation, edge-guided
interpolation, and texture enhancement. We show a schematic of our proposed method  in Figure \ref{fig:schematic}. Details of each step are given in the following.

\subsection{Edge Map Computation}

The edges in an image represent the boundaries between regions with significant changes in intensity (\cite{canny1986edge, marr1980theory}). 
The strength of an edge is determined by the magnitude of intensity change, while its direction indicates the orientation of the boundary.
%
We apply the Sobel operator \cite{canny1986edge} in \eqref{eq:Sobel} to compute the edge map of the input image.
%
\begin{equation}
    S_x = \begin{bmatrix} 
    -1 & 0 & 1 \\ 
    -2 & 0 & 2 \\ 
    -1 & 0 & 1 
    \end{bmatrix}, \quad
    S_y = \begin{bmatrix} 
    -1 & -2 & -1 \\ 
    0 & 0 & 0 \\ 
    1 & 2 & 1 
    \end{bmatrix},
    \label{eq:Sobel}
\end{equation}
where $S_x$ and $S_y$ are the Sobel kernels for the horizontal and vertical directions respectively. 
The resultant gradients $G_x$ and $G_y$ representing the rate of change of intensity in the horizontal and vertical directions as follows
\begin{equation}
    G_x = \frac{\partial I}{\partial x}, \quad G_y = \frac{\partial I}{\partial y}.
\end{equation}
Further,
\begin{equation}
    G_x(x, y) = \sum_{i=-1}^{1} \sum_{j=-1}^{1} I(x+i, y+j) \cdot S_x(i,j),
\end{equation}
and
\begin{equation}
    G_y(x, y) = \sum_{i=-1}^{1} \sum_{j=-1}^{1} I(x+i, y+j) \cdot S_y(i,j)
\end{equation}
The edge magnitude at each pixel $(x, y)$ is given by:
\begin{equation}
    E(x, y) = \sqrt{G_x(x, y)^2 + G_y(x, y)^2}.
\end{equation}
%
Next we normalize the edge map $E(x, y)$ as follows:
\begin{equation}
    I_{\text{E}}(x, y) = \frac{E(x, y) - E_{\text{min}}}{E_{\text{max}} - E_{\text{min}}}.
\end{equation}
where $E_{\text{min}}$ and $E_{\text{max}}$ are the minimum and maximum values of the edge map $E(x, y)$, respectively \cite{canny1986edge, marr1980theory}. This normalization ensures that the edge map is in a consistent range for further processing.
By applying the Sobel operator across the entire image, we obtain the edge map which highlights the regions with significant intensity transitions.

\begin{figure*}[htbp]
    \centering

    \subfloat[Edge map. ($I_E$).]{\includegraphics[height=38mm]{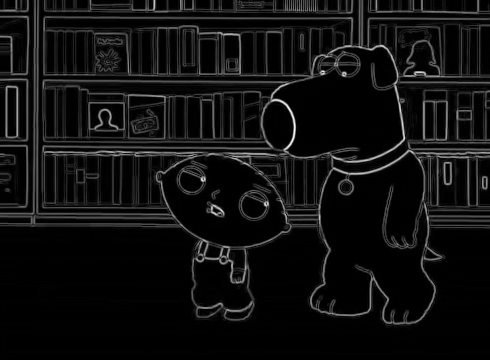}}\hfill
    \subfloat[Blurred image ($I_{\text{blur}}$).]{\includegraphics[height=38mm]{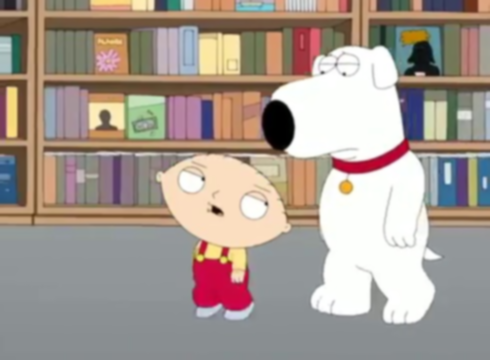}}\hfill
    \subfloat[Sharpened image ($I_s$).]{\includegraphics[height=38mm]{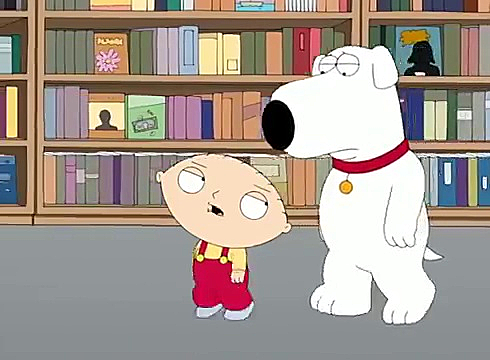}}\hfill
    \subfloat[Interpolated image ($I_B$).]{\includegraphics[height=38mm]{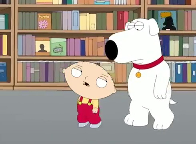}}\hfill
    \subfloat[Texture map ($I_T$).]{\includegraphics[height=38mm]{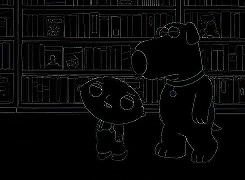}}\hfill
    \subfloat[Output image ($I_D$).]{\includegraphics[height=38mm]{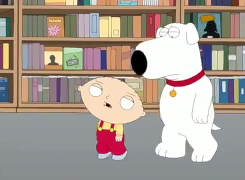}}

    \caption{Intermediate outputs of the structure-aware image downscaling algorithm. $I_E$ denotes the edge map, $I_{\text{blur}}$ the Gaussian-blurred image, and $I_s$ the sharpened image. $I_B$ is the bicubic-interpolated image, and $T$ is the texture map derived using a Laplacian operator. These components are fused to generate the final downscaled image $I_D$, preserving both structural and textural details.}
    \label{fig:structure_pipeline_stages}
\end{figure*}
\subsection{Edge-Guided Interpolation for Downscaling}

In this section, we introduce an enhanced downscaling technique that integrates edge information into the bicubic interpolation process to improve the preservation of structural integrity.  
Unlike standard bicubic interpolation \cite{keys1981cubic}, we perform edge-guided interpolation \cite{xu2009edge} to incorporate strong edge information into the interpolation process. This method adjusts the contribution of the bicubic interpolated image $I_{\text{B}}$ using the edge map $I_{\text{E}}$ as a weighting factor.

We first downscale the input image $I$ using standard bicubic interpolation, which estimates pixel values based on the surrounding $4 \times 4$ neighborhood of pixels. Suppose the bicubic interpolated output image is given by $I_B$.
Next, we compute the blurred image $I_{\text{blur}}$ by applying a Gaussian blur (or an equivalent smoothing technique) to obtain:
\begin{equation}
    I_{\text{blur}}(x, y) = \text{GaussianBlur}(I(x, y), \sigma),
\end{equation}
where $\sigma$ is the standard deviation for the Gaussian kernel. This operation smooths out the image, resulting in edges being less sharp.
To obtain a sharpened image $I_{\text{s}}$, we apply an unsharp mask technique, which enhances the high-frequency components of the image:
\begin{equation}
    I_{\text{s}}(x, y) = I(x, y) + \gamma \cdot (I(x, y) - I_{\text{blur}}(x, y)),
\end{equation}
where $\gamma$ is a parameter that controls the amount of sharpening.
Next, we construct an intermediate downscaled image $I'$ as a weighted combination of the bicubic interpolated image and the edge map $I_E$:
\begin{equation}
    I'(x, y) = (1 - I_{\text{E}}(x, y)) \odot I_{\text{B}}(x, y) + I_{\text{E}}(x, y) \odot I_{\text{s}}(x, y),
\end{equation}
where $I_{\text{E}}$ and $I_{\text{s}}$ are bicubic interpolated downscaled by factor of $d$.
The edge map $I_{\text{E}}(x, y)$ ranges from 0 to 1, with 1 indicating strong edges and 0 indicating smooth regions. The notation $\odot$ represents element-wise multiplication. This weight adjusts the influence of the sharp image and the bicubic image, ensuring that edges are better preserved.

By combining the bicubic interpolation with edge-guided enhancement \cite{xu2009edge}, this technique produces a sharper, more visually appealing downscaled image, especially in regions with significant edge content, such as text, architectural structures, or natural scenes with high-frequency details.

\begin{algorithm}
\caption{Structure-Aware Image Downscaling}
\begin{algorithmic}[1]
\State \textbf{Input:} High-resolution image $I \in \mathbb{R}^{M \times N}$; 
\Statex \hphantom{\textbf{Input:}} Downscaling factor $d$.
\State \textbf{Output:} Downscaled image $I_D \in \mathbb{R}^{m \times n}$, \Statex \hphantom{\textbf{Output:}}where $m = M/d$, $n = N/d$.
\Statex

\State \textbf{Edge Map Computation}
\vspace{0.2em}
\State $G_x = I * S_x$.  \Comment{convolution}
\State $G_y = I * S_y$. 
\State $E(x,y) = \sqrt{G_x(x,y)^2 + G_y(x,y)^2}$. 
\State $I_E(x,y) = \dfrac{E(x,y) - \min(E)}{\max(E) - \min(E)}$. 
\Statex

\State \textbf{Edge-Guided Interpolation}
\vspace{0.2em}
\State $I_B = \text{Bicubic}(I, d)$. \Comment{interpolation by factor $d$}
\State $I_E^{\text{down}} = \text{Bicubic}(I_E, d)$.
\State $I_{\text{blur}} = \text{GaussianBlur}(I)$. 
\State $I_s(x,y) = I(x,y) + \gamma \cdot (I(x,y) - I_{\text{blur}}(x,y))$. 
\State $I_s^{\text{down}} = \text{Bicubic}(I_s, d)$.
\State $I'_1(x,y) = (1 - I_E^{\text{down}}(x,y)) \odot I_B(x,y)$. 
\State $I'_2(x,y) =  I_E^{\text{down}}(x,y) \odot I^{\text{down}}_s(x,y)$.
\State $I'(x,y) = I'_1 + I'_2$. 
\Statex

\State \textbf{Texture Enhancement}
\vspace{0.2em}
\State $I_T(x,y) = \text{Laplacian}(I)$. 
\State $I_T^{\text{down}} = \text{Bicubic}(I_T, d)$.
\State $\lambda^{\text{down}} = \alpha  I_E^{\text{down}} + \beta$. 
\State $I_D(x,y) \gets I'(x,y) + \lambda^{\text{down}}(x,y) \odot I_T^{\text{down}}(x,y)$. 
\end{algorithmic}
\end{algorithm}

\subsection{Local Texture based Expansion}

This step of the algorithm consists of two major steps: computing the edge map and texture-guided local expansion.

\subsubsection{Texture Map Computation}

Texture plays a crucial role in visual perception by capturing fine structures, patterns, and high-frequency components. However, traditional downscaling methods often suppress these details, leading to overly smooth outputs. To mitigate this, preserving texture information during downscaling is essential.

Various techniques, such as frequency-domain filtering, wavelet methods, and deep learning, have been proposed for texture preservation. However, these approaches can be computationally expensive and may not generalize well. In this work, we use a Laplacian-based texture extraction method, which is computationally efficient and effective for identifying high-frequency components critical for texture preservation.

The Laplacian operator is as follows a second-order derivative filter that highlights regions of rapid intensity change, emphasizing fine details such as edges and texture patterns: 
\begin{equation}
    \frac{\partial^2 I(x, y)}{\partial x^2} + \frac{\partial^2 I(x, y)}{\partial y^2}.
\end{equation}
This derivative is approximated by convolution with the Laplacian kernel:
\begin{equation}
    L = \begin{bmatrix}
    0 & 1 & 0 \\
    1 & -4 & 1 \\
    0 & 1 & 0
    \end{bmatrix}.
\end{equation}
By convolving the image \( I(x, y) \) with this kernel, we obtain the texture map \( I_{\text{T}}(x, y) \):
\begin{equation}
    I_{\text{T}}(x, y) = \sum_{i=-1}^{1} \sum_{j=-1}^{1} I(x+i, y+j) \cdot L(i, j)
\end{equation}
This texture map is normalized for consistent scaling and used in the final image reconstruction. The Laplacian kernel enhances intensity transitions, capturing edges and textures, ensuring  high-quality downscaled images.
The texture map \( I_{\text{T}}(x, y) \) plays a key role in restoring fine details, resulting in more accurate and visually appealing downscaled images. A visual representation of the texture map is shown in Fig. \ref{fig:structure_pipeline_stages}.
\subsubsection{Texture-Guided Local Expansion}

To ensure texture preservation in the final downscaled image, we introduce a texture-guided local expansion step. Traditional downscaling methods typically rely solely on interpolated pixel values, which may lack fine details. By incorporating the texture map into the reconstruction process, we enhance texture fidelity while maintaining the overall structure of the image.

In this approach, we refine the interpolated image $I'$ by adding the texture map $I_{\text{T}}$ with an adaptive weight parameter $\lambda$, which controls the contribution of texture information:
\begin{equation}
I_{\text{D}}(x, y) = I'(x, y) + \lambda I_{\text{T}}(x, y)
\end{equation}
The weight parameter $\lambda$ is dynamically adjusted based on the local edge strength $I_{\text{E}}(x, y)$, which helps prevent over-amplification of texture in smoother regions. The adaptive weight function is defined as:
\begin{equation}
\lambda(x, y) = \alpha I_{\text{E}}(x, y) + \beta
\end{equation}
where $\alpha$ and $\beta$ are scaling factors that control the influence of texture. Higher values of $I_{\text{E}}(x, y)$, corresponding to stronger edges, result in more pronounced texture restoration, while smoother regions receive minimal enhancement. This dynamic adjustment ensures that texture restoration is context-sensitive, leading to an image with both enhanced detail and smooth areas where necessary.

This approach ensures that textures are meaningfully restored in regions with significant details, while smooth regions are left untouched, avoiding the introduction of artifacts. The result is a perceptually superior downscaled image that maintains both the fine-grained texture and the overall structure, enhancing the quality of the downscaling process.

\begin{table*}
    \centering
        \caption{Quantitative results (PSNR / SSIM) for  scaling factor $d = 4$ on different datasets. The \textbf{best} results are marked in red.}
    \begin{tabular}{|c|p{2.5cm}|p{2.2cm}|p{2.2cm}|p{2.2cm}|p{2.2cm}|}
        \hline
        {Downscaling Method} & Reference  & DIV2K Dataset & BSD100 Dataset & Urban100 Dataset & RealSR Dataset   \\
         \hline
        Bicubic & IEEE TASSP 1961  & 30.47 / 0.942 & 26.34 / 0.847 & 23.46 / 0.826 & 31.11 / 0.918  \\ 
        Lanczos & JSIAM 1964  & 30.35 / 0.950 & 26.72 / 0.849 & 23.69 / 0.831 & 31.29 / 0.913  
        \\
        Content-adaptive & ACM TOG 2013  & 29.04 / 0.950 & 27.11 / 0.851 & 23.93 / 0.832 & 31.32 / 0.923  
        \\
        DPID & ACM TOG 2016  & 32.98 / 0.950 & 39.57 / 0.989 & 36.20 / 0.986 & 30.30 / 0.869  
        \\
        IDCL & JVIS 2023  & 36.37 / 0.98 & 36.27 / 0.979 & 32.12 / 0.971 & 28.10 / 0.876 \\
        ADL & IEEE TPAMI 2022  & 33.05 / 0.931 & 29.12 / 0.873 & 26.81 / 0.867 & 34.50 / 0.965 
        \\
        SDFlow & IEEE TPAMI 2024  & 33.62 / 0.946 & 29.51 / 0.892 & 26.95 / 0.872 & 34.85 / 0.966 
        \\
        SAID & Proposed  & \textbf{39.07} / \textbf{0.982} & \textbf{37.67} / \textbf{0.994} & \textbf{34.17} / \textbf{0.988} & \textbf{38.71} / \textbf{0.996} 
        \\
        \hline
    \end{tabular}
    \label{tab:comparison}
\end{table*}

\section{Experimental Results}
\label{sec:results}
%
%
\subsection{Datasets}
We conducted experiments on a diverse and extensive set of publicly available high-resolution image datasets: DIV2K \cite{agustsson2017ntire}, BSD100 \cite{martin2001database}, Urban100 \cite{huang2015single}, and RealSR \cite{cai2019toward}. The DIV2K (Diverse 2K) dataset, introduced in the NTIRE 2017 Challenge on Single Image Super-Resolution \cite{agustsson2017ntire}, is a widely recognized benchmark for image restoration and downscaling tasks. 
This DIV2K  dataset includes ground-truth downscaled versions for scale factor 4, which were used for objective evaluation at those scales.
To further validate the generalization capability of our method, we evaluated performance on BSD100 \cite{martin2001database}, and Urban100 \cite{huang2015single}, and RealSR \cite{cai2019toward} datasets. We conducted experiments for integer scale factors \(s = \{2, 4, 8, 16\}\).
We note that Urban100 dataset \cite{huang2015single}, which consists of structurally complex scenes with repeated patterns—providing a challenging testbed to assess the preservation of fine geometric details. Lastly, a curated subset of high-resolution natural images from the ImageNet dataset \cite{deng2009imagenet} was used to evaluate performance on semantically rich and diverse real-world content.
The collection of four total datasets ensures a robust and holistic evaluation across multiple image categories and complexity levels, and across both fixed and arbitrary downscaling factors.







\subsection{State-of-the-art Comparison of Downscaling Performance}

We evaluated our algorithm across multiple high-resolution image datasets to comprehensively assess its downscaling performance.. The primary objective was to verify the algorithm’s ability to preserve structure, texture, and perceptual quality in a wide class of images.
For a given scale factor, the algorithm was applied to high-resolution input images, and the resulting downscaled outputs were compared against either available ground-truth references or standard interpolation-based baselines.
Performance was quantitatively assessed using two image quality metrics: Peak Signal-to-Noise Ratio (PSNR) and Structural Similarity Index (SSIM) \cite{wang2004image}. In addition, qualitative visual comparisons were performed to subjectively evaluate perceptual fidelity and structural preservation on all evaluated scales.

For quantitative evaluation, we compare our proposed SAID method with recent classical techniques for image downscaling: bicubic interpolation \cite{keys1981cubic}, Lanczos resampling \cite{lanczos1964interpolation}, content-aware downscaling \cite{kopf2013content}, detail preserving image downscaling (DPID) \cite{weber2016rapid}, and image downscaling via co-occurrence learning (IDCL) \cite{ghosh2023image}. We also compare with state-of-the-art deep learning methods\footnote{For deep learning methods, we have used pre-trained model publicly provided by the authors.} for image downscaling: adaptive downsampling models (ADL) \cite{son2022toward} and SDFlow \cite{sun2023sdflow}.
The results are summarized in Table~\ref{tab:comparison}, with the best values in each row highlighted in bold. All these reported results are for the downscaling factor $d = 4$.

For DIV2K dataset \cite{agustsson2017ntire}, our method SAID achieves the highest PSNR \textbf{39.07 dB} and SSIM \textbf{0.982}, where the latest SDFlow method (deep learning) gives PSNR 33.62 dB and SSIM 0.946. The enriched collection of natural images in the DIV2K dataset confirms that SAID outperforms in producing state-of-the-art downscaling of natural images. 
The structurally complex scenes with repeated patterns in the images in the BSD100 \cite{martin2001database} and Urban100 \cite{huang2015single} datasets make it a challenging task to preserve the fine geometric details in the downscaled images. This is reflected in the relatively lower values of PSNR and SSIM in Table~\ref{tab:comparison} for the BSD100 \cite{martin2001database} and Urban100 \cite{huang2015single} datasets. However, our method SAID still offers the best performance in terms of both PSNR and SSIM metrics. Finally, for the RealSR dataset, SAID outperforms by achieving the highest PSNR \textbf{37.71 dB} and SSIM \textbf{0.996} values. The very high SSIM value for the RealSR dataset confirms that SAID could successfully retain strong structural information in the downscaled image. In summary, the experimental results in Table~\ref{tab:comparison} demonstrate the effectiveness of our SAID method to retain both pixel-level and perceptual quality for a wide variety of image datasets.

\begin{figure*}
    \centering

    \subfloat[Input.]{\includegraphics[width=0.25\textwidth]{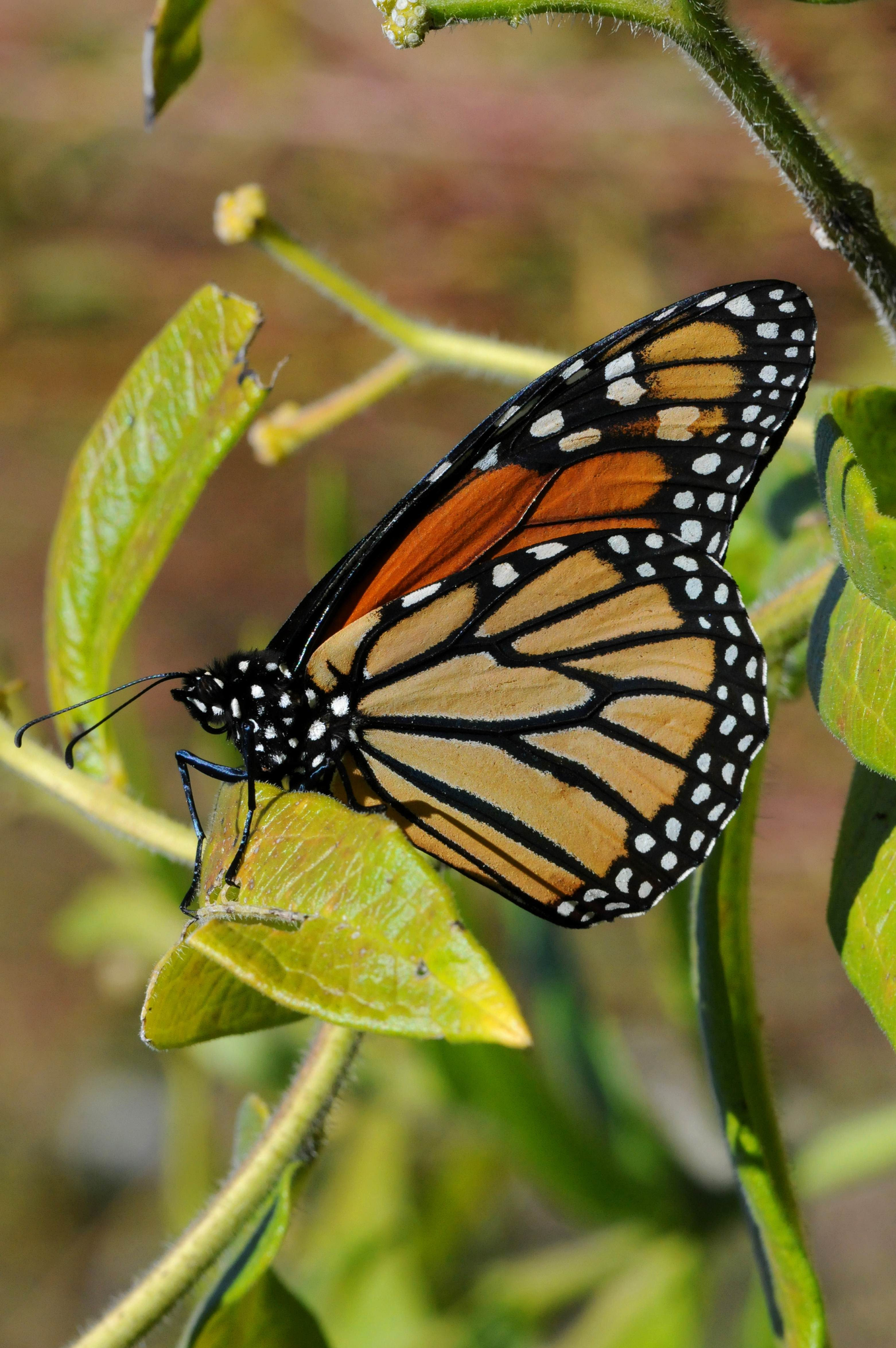}} \hspace{1em}
    \subfloat[Bicubic \cite{keys1981cubic} \textcolor{red}{(32.42, 96.5)}.]{\includegraphics[width=0.25\textwidth]{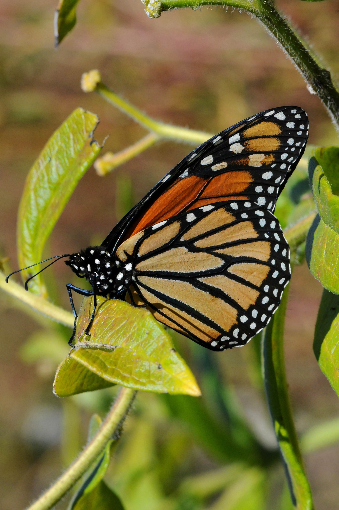}}\hspace{1em}
    \subfloat[SDFlow \cite{sun2023sdflow} \textcolor{red}{(30.4, 93.6)}.]{\includegraphics[width=0.25\textwidth]{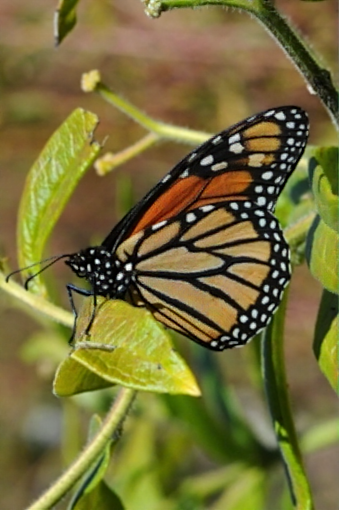}}
    \\
%
    \subfloat[DPID \cite{weber2016rapid} \textcolor{red}{(37.16, 98.6)}.]{\includegraphics[width=0.25\textwidth]{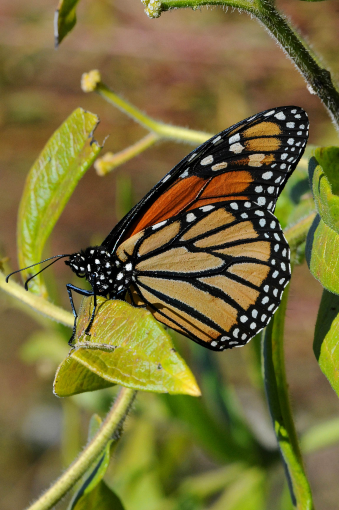}} \hspace{1em}
    \subfloat[IDCL \cite{ghosh2023image} \ \textcolor{red}{(44.83, 99.6)}.]{\includegraphics[width=0.25\textwidth]{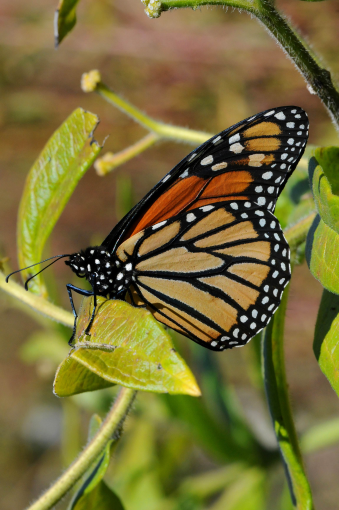}} \hspace{1em}
    \subfloat[SAID \ \textcolor{red}{(45.5, 99.8)}.]{\includegraphics[width=0.25\textwidth]{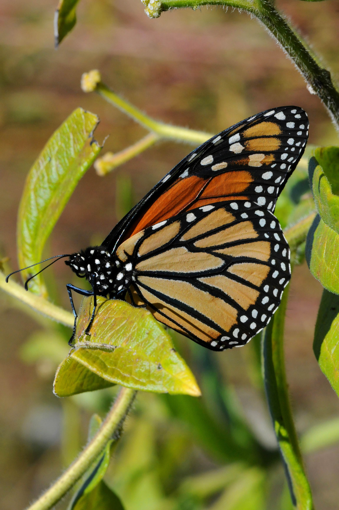}}
    \caption{Visual comparison of cartoon image with different image downscaling techniques: bicubic \cite{keys1981cubic}, SDFlow\cite{sun2023sdflow}, DPID \cite{weber2016rapid}, IDCL \cite{ghosh2023image} and our method SAID for the downscaling factor of $4$. and we can notice lesser artifacts and sharp fine details retention.}
    \label{fig:butterfly_compare_4}
\end{figure*}
\begin{figure*}
    \centering
    \subfloat[Bicubic \cite{keys1981cubic}.]{\includegraphics[width=0.23\textwidth]{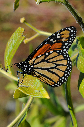}}\hspace{1em}
        \subfloat[DPID.]{\includegraphics[width=0.23\textwidth]{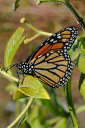}} \hspace{1em}
    \subfloat[SDFlow \cite{sun2023sdflow}.]{\includegraphics[width=0.23\textwidth]{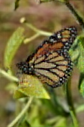}} \hspace{1em}
    \subfloat[SAID.]{\includegraphics[width=0.23\textwidth]{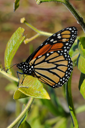}}
    \caption{Visual comparison of butterfly image with different image downscaling techniques: bicubic \cite{keys1981cubic}, SDFlow\cite{sun2023sdflow}, DPID \cite{weber2016rapid}, IDCL \cite{ghosh2023image} and our method SAID for the \textbf{downscaling factor of $16$}. We notice lesser artifacts and sharp fine details retention.}
    \label{fig:butterfly_compare_16}
\end{figure*}

\begin{figure*}[htbp]
    \centering

    \subfloat[Input.]{\includegraphics[width=0.3\textwidth]{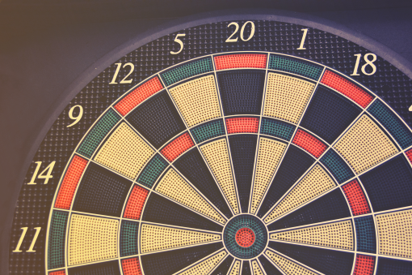}}\hfill
    \subfloat[Bicubic \cite{keys1981cubic}.]{\includegraphics[width=0.3\textwidth]{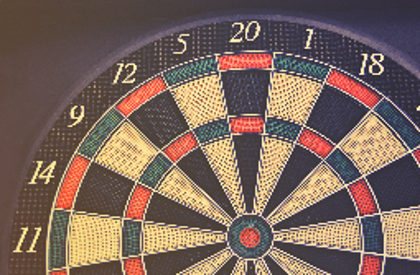}}\hfill
    \subfloat[Content \cite{kopf2013content}.]{\includegraphics[width=0.3\textwidth]{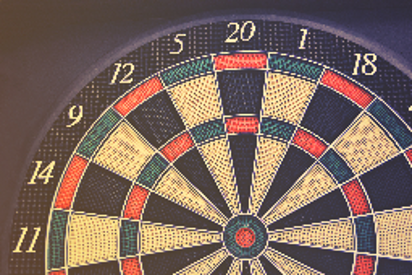}}
    \vspace{1em}

    \subfloat[DPID \cite{weber2016rapid}.]{\includegraphics[width=0.3\textwidth]{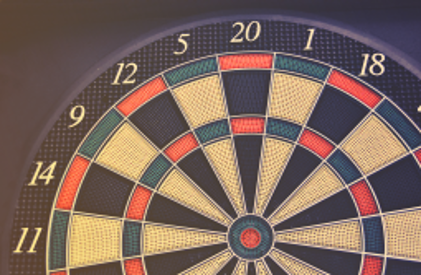}}\hfill
    \subfloat[IDCL \cite{ghosh2023image}.]{\includegraphics[width=0.3\textwidth]{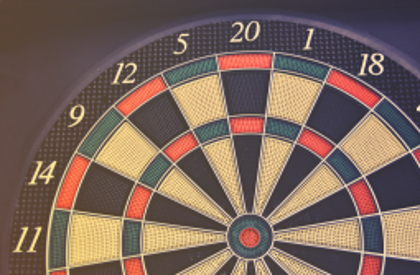}}\hfill
    \subfloat[SAID (ours).]{\includegraphics[width=0.3\textwidth]{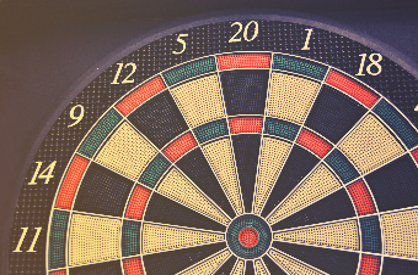}} \\
        \subfloat[Input (zoomed-in).]{\includegraphics[width=0.22\textwidth]{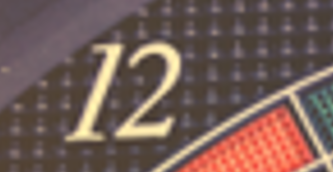}} \hspace{2em}
    \subfloat[Bicubic (zoomed-in).]{\includegraphics[width=0.22\textwidth]{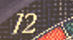}} \hspace{2em}
    \subfloat[Content  (zoomed-in).]{\includegraphics[width=0.22\textwidth]{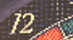}} \\
    \subfloat[DPID (zoomed-in).]{\includegraphics[width=0.22\textwidth]{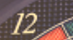}} \hspace{2em}
    \subfloat[IDCL (zoomed-in).]{\includegraphics[width=0.22\textwidth]{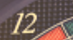}} \hspace{2em}
    \subfloat[ SAID (zoomed-in).]{\includegraphics[width=0.22\textwidth]{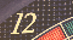}}
    \caption{Visual comparison of a downscaled texture image  by  different  downscaling techniques: bicubic \cite{keys1981cubic}, content-daptive \cite{kopf2013content}, DPID \cite{weber2016rapid}, IDCL \cite{ghosh2023image} and our method SAID for downscaling factor of $8$. In zoomed-in version of the number 12 of the dartboard, we see our method (SAID) efficiently retains the sharpness and texture pattern, such as the dots around the number and edges of number, resulting in superior visual quality as compared to other approaches.}
    \label{fig:dartboard_comparison}
\end{figure*}


\subsection{Visual Comparisons on Different Image Types}

In this section, we perform a detailed qualitative analysis by evaluating our method across four distinct image categories-natural scenes, texture-dense images, text-rich content, and cartoon illustrations. This allows us to assess how well SAID generalizes across visually diverse contexts and use cases.

\subsubsection{\textbf{Result on Natural Image}}
Natural images often contain complex structures, intricate patterns, and fine textures—such as the fur of an animal, leaves on trees, or skin tones—all of which contribute to the perceptual realism of the image. Downscaling such images is particularly challenging because it requires preserving a wide range of spatial frequencies without introducing blurring or artificial enhancements. In Figure \ref{fig:butterfly_compare_4}, we show a visual result of donwscaling a natural image, referred as \textit{butterfly}, by our method SAID and other recent methods. 
%
We found that high-frequency components, such as edges and textures, tend to blur essential details by most of the compared methods. 
%
We also see some sort of color distortion and undesirable halo artifacts and edge ringing, especially around the high-contrast boundary lines.. 
The highest values of (PSNR, SSIM) validate the superiority of our SAID method in maintaining structural content and preserving the best possible similarity to the ground-truth image. By incorporating edge-guided interpolation and structural awareness during downsampling, SAID avoids the typical trade-off between detail retention and natural appearance, yielding outputs that are perceptually close to the original high-resolution image even at a $4 \times$ downscaling factor.

To further study the effect of large scale downscaling, we show the results for scaling factor $d = 16$ and compare with the output of bicubic, DPID, and SDFlow in Figure~\ref{fig:butterfly_compare_16}. The downscaled output image by the bicubic method exhibits severe blocking artifacts. The DPID method results in blocking artifact and unwanted blurring causing loss of sharp edges. The output of SDFlow method suffers from a severe blurring effect, resulting in a significant decline in visual quality. Our SAID method is cabable of producing a better visual quality among all the methods in Figure~\ref{fig:butterfly_compare_16}.

\begin{figure*}[htbp]
    \centering

    \subfloat[Input.]{\includegraphics[width=0.3\textwidth]{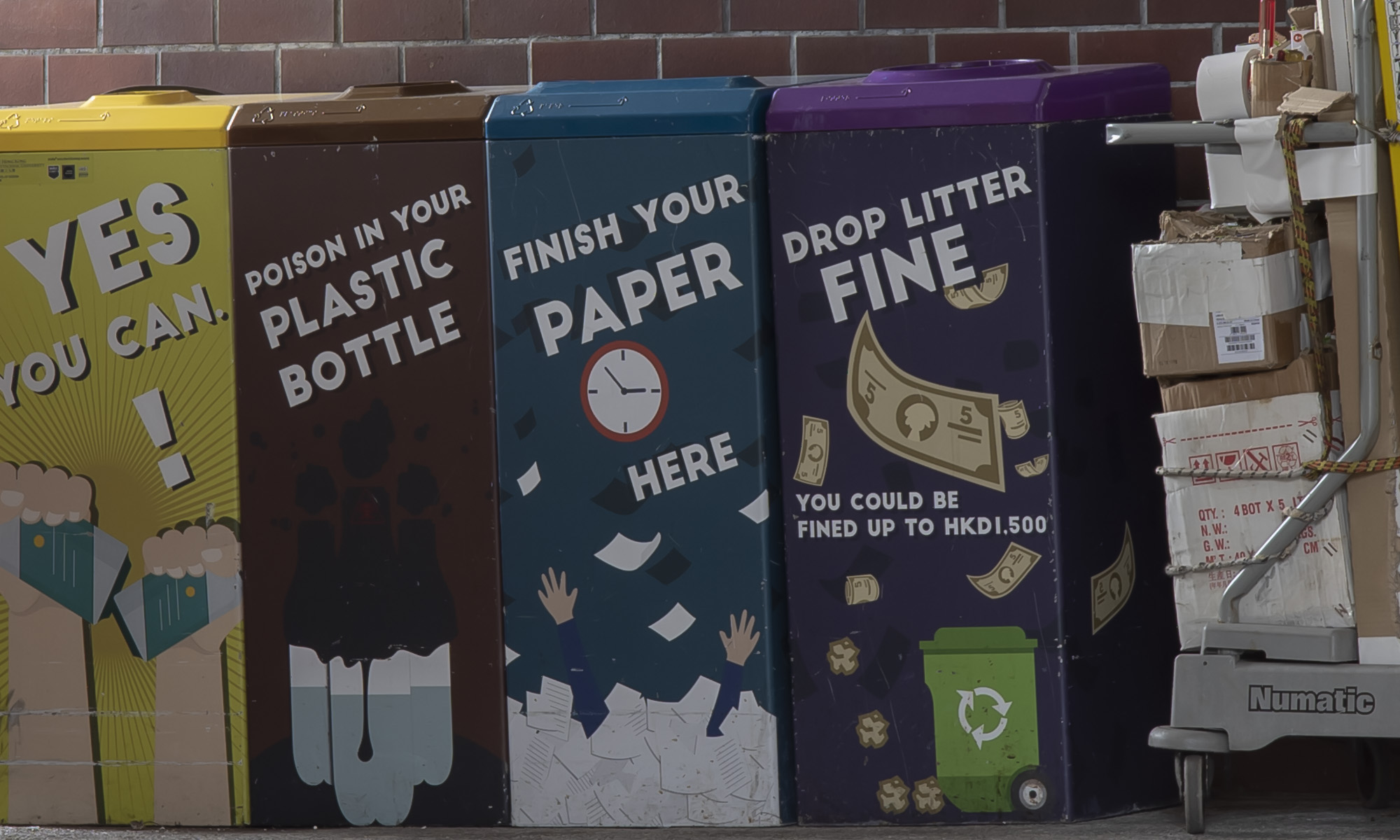}}\hfill
    \subfloat[Bicubic.]{\includegraphics[width=0.3\textwidth]{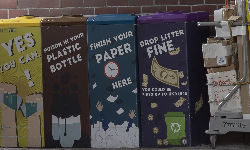}}\hfill
    \subfloat[SDflow.]{\includegraphics[width=0.3\textwidth]{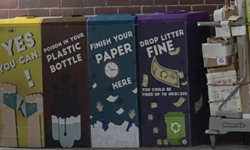}}

    \vspace{1em}

    \subfloat[DPID.]{\includegraphics[width=0.3\textwidth]{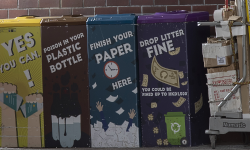}}\hfill
    \subfloat[IDCL.]{\includegraphics[width=0.3\textwidth]{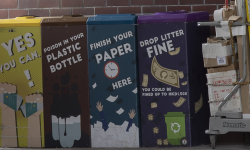}}\hfill
    \subfloat[SAID.]{\includegraphics[width=0.3\textwidth]{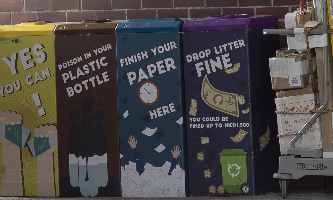}} \\
    \subfloat[Input (zoomed-in).]{\includegraphics[width=0.28\textwidth]{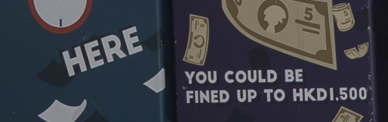}} \hspace{2em}
    \subfloat[Bicubic (zoomed-in).]{\includegraphics[width=0.25\textwidth]{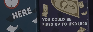}} \hspace{2em}
    \subfloat[Sdflow (zoomed-in).]{\includegraphics[width=0.25\textwidth]{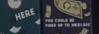}} \\
    \subfloat[DPID (zoomed-in).]{\includegraphics[width=0.25\textwidth]{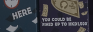}} \hspace{2em}
    \subfloat[IDCL (zoomed-in).]{\includegraphics[width=0.25\textwidth]{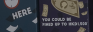}} \hspace{2em}
    \subfloat[SAID (zoomed-in).]{\includegraphics[width=0.25\textwidth]{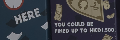}}
    \caption{Visual comparison with different image downscaling techniques: bicubic \cite{keys1981cubic},lancoz\cite{lanczos1964interpolation}, DPID \cite{weber2016rapid}, IDCL \cite{ghosh2023image} and our method SAID for the downscaling factor of $8$.In zoomed in version of the text written in box 3 and 4 we can see our method (SAID) efficiently retains the text written with better sharpness, such as clarity in smaller text, resulting in better detail retention without over smoothing the image. A zoomed-in views are also shown here. Notice that SAID reconstructs the fine text\textbf{ “YOU COULD BE FINED UP TO HKD1.500”} with superior legibility and crispness than all other methods compared.}    
    \label{fig:advt}
\end{figure*}
\subsubsection{\textbf{Result on Texture Image}}
Images that are texture-dense—such as those depicting fabrics, tiles, or complex geometric patterns like a dartboard—require a downscaling method capable of maintaining repeating structures, directional patterns, and sharp edges. 
In Figure \ref{fig:dartboard_comparison}, the dartboard image evaluated in our study includes radial lines, concentric circles, and numerical digits, making it an ideal benchmark for texture preservation. In this experiment, bicubic \cite{keys1981cubic} and Lanczos \cite{lanczos1964interpolation} methods tend to oversmooth such repetitive patterns, leading to blurred edges and degraded clarity in critical regions, such as the digits or the central bullseye. Notice in the zoomed view in Figure \ref{fig:dartboard_comparison} that 
{DPID \cite{weber2016rapid} improves contrast but does not preserve the structural alignment of patterns, therefore resulting in blocky textures and halo effects.} IDCL \cite{ghosh2023image}, while capable of edge-aware interpolation, tends to smooth textures too aggressively, especially in areas with low texture contrast, thereby eroding structural consistency. In contrast, it is evident in Figure \ref{fig:dartboard_comparison}  that SAID reconstructs the digit outlines and circular patterns with high accuracy, maintaining symmetry and alignment. The numerical elements around the dartboard remain legible, and the color bands are preserved without ringing artifacts. This indicates that SAID is particularly well-suited for retaining structural regularity and textural sharpness which are critical in applications  such as technical illustrations, patterned designs, and texture analysis.

\subsubsection{\textbf{Result on Document Image}}
Textual content presents one of the most stringent tests for any downscaling algorithm. Even minor distortions in character edges or spacing can dramatically affect readability. In images containing signage or dense textual content, clarity and sharpness are non-negotiable. 
We perform an experiment of downscaling an image with important signage content in it and the results are shown in Figure \ref{fig:advt}. 
We see that  bicubic \cite{keys1981cubic} and Lanczos \cite{lanczos1964interpolation}  produce soft edges and blurred characters, which is particularly problematic for small fonts or fine print. Lack of sharpness reduces legibility and impairs usability in practical scenarios such as document thumbnails, street signs, and user interface elements.
Notice in Figure~\ref{fig:advt}  that DPID \cite{weber2016rapid} introduces artificial contrast that can distort thin strokes and make characters look uneven or jagged. IDCL \cite{ghosh2023image}, while preserving edge structure, sometimes over-smooths strokes or alters character geometry slightly, resulting in a loss of typographic fidelity. SDFlow fails to retain useful text information in the downscaled output. Our SAID method outperforms these alternatives by precisely reconstructing text contours, maintaining semantic boundaries, and suppressing unnecessary smoothing. For example, the reconstructed text \textbf{“YOU COULD BE FINED UP TO HKD1.500”} in the zoomed-in view in Figure \ref{fig:advt} retains its typographic form with high clarity and alignment. It is hardly possible to make out the text in the zoomed-in view of the output image using SDFlow. Notice that the best possible readability of the text sentence is achieved in the zoom-in views of the downscaled output images in Figure~\ref{fig:advt}.
SAID achieves this through edge-informed guidance and preservation of local structure, making it particularly effective in scenarios where text content must remain legible and visually consistent across resolution changes.

\section{Conclusion}
\label{sec:conclusion}
We  introduced a new algorithm for image downscaling. The core idea combines edge-guided interpolation and texture expansion to preserve fine details.
Our method offered competitive performance with reference to the existing methods. It could successfully preserve structure of an image while minimizing artifacts. 
The superior performance of our approach was obtained by the integration of an edge-guided interpolation mechanism with a local expansion strategy. 
The proposed method exhibited enhanced edge preservation by leveraging edge-aware processing. Therefore, our method SAID could effectively accentuated crucial boundaries, leading to sharper and more accurate edge representation.   Further, our method SAID could be extended for non-integer scaling factors, overcoming limitations of some of traditional techniques, making it suitable for real-world applications.
%
%
Future work can explore extending our method to video downscaling, where temporal consistency must be preserved alongside edge-aware interpolation. The proposed efficient downscaling method could be used in applications such as medical imaging and remote sensing. 

\bibliographystyle{IEEEbib}
\bibliography{strings}
\begin{IEEEbiography}[{\includegraphics[width=1in,height=1.25in,clip,keepaspectratio]{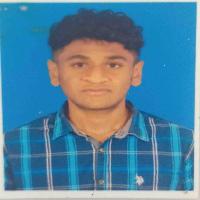}}]{G. B. Kevin Arjun} received the M.Tech. degree in Signal Processing and Machine Learning from the Indian Institute of Technology Kharagpur, India. He is currently working on advanced topics in image processing and deep learning, with a focus on image downscaling, generative AI, and visual representation learning. His research interests include image compression, image downscaling and super resolution.  
\end{IEEEbiography}

\begin{IEEEbiography}[{\includegraphics[width=1in,height=1.25in,clip,keepaspectratio]{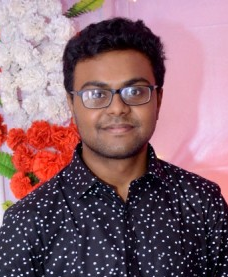}}]{Suvrojit Mitra}  received his B.Tech. degree in Electrical Engineering from the University Institute of Technology, The University of Burdwan, West Bengal, India, in 2020, and the M.Tech. degree in Electronics and Communication Engineering with a specialization in Signal Processing from Dr B R Ambedkar National Institute of Technology, Jalandhar, Punjab, India, in 2024. He is currently pursuing the Ph.D. degree in Electrical Engineering at Indian Institute of Technology Kharagpur. His research interests include image processing, particularly in image enhancement and denoising and image and video downscaling.
\end{IEEEbiography}

\begin{IEEEbiography}[{\includegraphics[width=1in,height=1.25in,clip,keepaspectratio]{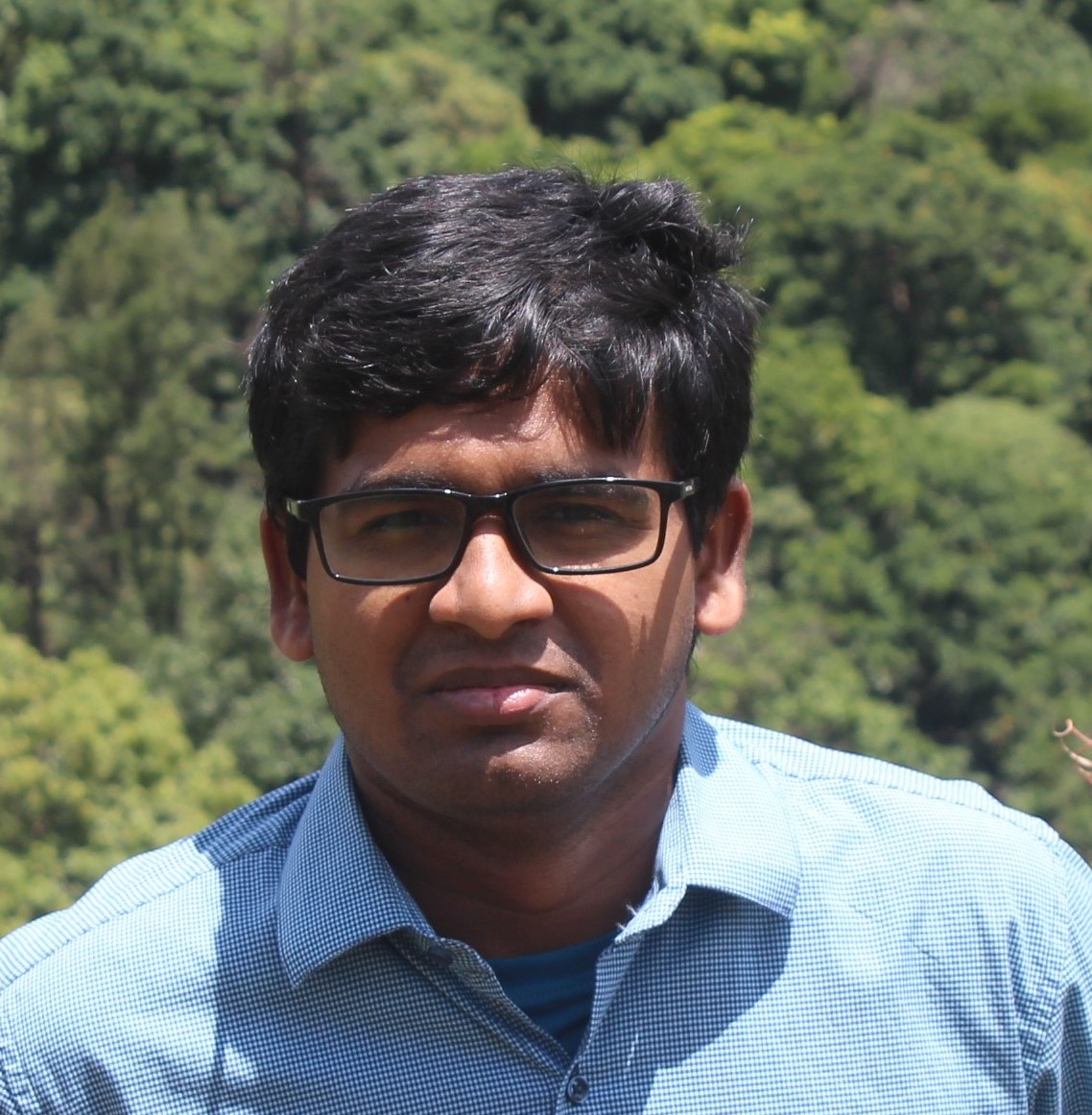}}]{Sanjay Ghosh} received the PhD degree in Electrical Engineering from the Indian Institute of Science in 2019. Cuurently, he is an Assistant Professor in the Department of Electrical Engineering at the Indian Institute of Technology Kharagpur, India. 
His broad research interests are in computational imaging, brain signal processing, and machine learning methods for neurological disorder analysis. He received Best Student Paper Award at IEEE Global Conference on Signal and Information Processing (GlobalSIP) 2018 and Silver Award at International Conference on Biomagnetism (BIOMAG) 2022. 
\end{IEEEbiography}

\end{document}